\documentclass[a4paper,fleqn,usenatbib]{mnras}

\usepackage{newtxtext,newtxmath}
\usepackage{ulem}

\usepackage[T1]{fontenc}
\usepackage{ae,aecompl}

\usepackage{graphicx}	
\usepackage{amsmath}	
\usepackage{amssymb}	

\pdfoutput=1

\usepackage{url}
\usepackage{amssymb,amsmath}
\usepackage{subfigure}
\usepackage{booktabs}

\usepackage{natbib}

\usepackage{color,hyperref}
\definecolor{linkcolor}{rgb}{0,0,0.5}
\hypersetup{colorlinks=true,linkcolor=linkcolor,citecolor=linkcolor,
            filecolor=linkcolor,urlcolor=linkcolor}
\usepackage{url}
\usepackage{subfigure}
\usepackage{booktabs}

\usepackage{natbib}

\newcommand{\msun}{${\rm M}_{\odot}$}

\newcommand{\mstar}{${\rm M}_{\star}$}
\newcommand{\mhalo}{${\rm M}_{\rm halo}$}
\newcommand{\lstar}{${\rm L}_{\star}$}
\newcommand{\muv}{${\rm M}_{\rm UV}$}
\newcommand{\mv}{${\rm M}_{\rm V}$}
\newcommand{\mmin}{${\rm M}_{\rm min}$}
\newcommand{\hst}{\textit{HST}}
\newcommand{\jwst}{\textit{JWST}}
\newcommand{\fesc}{${\rm f}_{\rm escape}$}

\title[Ultra-Faint Galaxies and Reionization]{Local Group Ultra-Faint Dwarf Galaxies in the Reionization Era}

\author[D. R. Weisz et al.]{
Daniel R. Weisz$^{1}$,
Michael Boylan-Kolchin$^{2}$\\
$^{1}$Department of Astronomy, University of California Berkeley, Berkeley, CA 94720, USA; dan.weisz@berkeley.edu\\
$^{2}$Department of Astronomy, The University of Texas at Austin, 2515 Speedway, Stop C1400, Austin, TX 78712-1205, USA\\
}

\pubyear{2017}

\begin{document}
\label{firstpage}
\pagerange{\pageref{firstpage}--\pageref{lastpage}}
\maketitle

\begin{abstract}
Motivated by the stellar fossil record of Local Group (LG) dwarf galaxies, we show that the star-forming ancestors of the faintest ultra-faint dwarf galaxies (UFDs; \mv\ $\sim -2$
  or \mstar\ $\sim 10^{2}$ at $z=0$) had ultra-violet (UV) luminosities of
  \muv\ $\sim -3$ to $-6$ during reionization ($z\sim6-10$).  The
  existence of such faint galaxies has substantial implications for early epochs of galaxy formation and reionization.  If the faint-end slopes of the UV luminosity functions
  (UVLFs) during reionization are steep ($\alpha\lesssim-2$) to \muv\ $\sim -3$, 
  then: (i) the ancestors of UFDs produced $>50$\% of UV flux from galaxies; (ii) galaxies can maintain reionization with escape fractions that are $>$2 times
  lower than currently-adopted values; (iii) direct HST and JWST observations
  may detect only $\sim10-50$\% of the UV light from galaxies;
  (iv) the cosmic star formation history increases by  $\gtrsim4-6$ at $z\gtrsim6$. 
 Significant flux from UFDs, and resultant tensions with LG dwarf galaxy
  counts, are reduced if the high-redshift UVLF turns over.
  Independent of the UVLF shape, the existence of a large population of UFDs \textit{requires} a non-zero luminosity function to \muv\ $\sim -3$ during reionization.
\end{abstract}




\section{Introduction}

Low-mass galaxies appear to play a central role in ionizing the neutral
intergalactic medium (IGM) in the early Universe.  While sources such as AGN and
X-ray binaries may contribute ionizing photons \citep[e.g.,][]{mcquinn2012,
  madau2015}, the consensus view is that a large population of low-mass galaxies
is required to explain the observed ionization fraction of the IGM and the
Thomson optical depth ($\tau_{\rm e}$) during reionization \citep[e.g.,][]{stark2016}.

However, the low-mass galaxy population associated with reionization has never
been directly detected.  The deepest available blank-field \textit{Hubble Space
  Telescope (HST)} observations only reach \muv($z\sim7)\sim-16$
\citep[e.g.,][]{finkelstein2015, bouwens2015}, but contemporary models of the
high-$z$ Universe \citep[e.g.,][]{kuhlen2012, robertson2013, robertson2015}
require the ultra-violet luminosity function (UVLF) to remain steep to
magnitudes as faint as \muv($z\sim7)=-10$ in order for galaxies to maintain
reionization.  Gravitational lensing and the exquisite sensitivity of the
\textit{James Webb Space Telescope (JWST)} promise to push observations further
down the galaxy UVLF (e.g., to \muv($z\sim7)\sim-14$), but the direct detection and
characterization of a significant number of intrinsically faint galaxies
(\muv\ $\sim -10$) in the reionization era appears beyond observational
capabilities for the foreseeable future.  Consequently, the contribution of
low-mass galaxies to reionization will continue to be estimated by extrapolating
the observed UVLF to galaxies that cannot be directly observed.

Concurrent with searches for faint galaxies at high redshifts, there has been a
renaissance in the discovery of extremely faint, low-mass galaxies in the Local
Group (LG; e.g., \citealt{willman2005, belokurov2006, belokurov2010, irwin2007,
  koposov2007, zucker2006, zucker2006b, kim2015, koposov2015, laevens2015,
  laevens2015b, martin2015, des2015}).  Deep, wide-field photometric surveys
(e.g., SDSS, DES) have identified dozens of faint galaxies surrounding the Milky
Way (MW) with luminosities as low as \mv\ $\sim-2$ (\mstar\ $\sim 10^2$ \msun), and
these detections likely represent only a fraction of the low-mass galaxy
population in the Local Group owing to various observational biases
\citep[e.g.,][]{koposov2008, tollerud2008, walsh2009}.  These so-called
`ultra-faint' dwarf galaxies (UFDs) host predominantly ancient ($>13$ Gyr),
extremely metal-poor ([Fe/H]$<-2$) stellar populations
\citep[e.g.,][]{frebel2015}. They are therefore consistent with being ``fossils
of reionization'' \citep[e.g.,][]{ricotti2005, okamoto2012, brown2014,
  weisz2014b}: UFDs were star-forming galaxies in the early Universe until UV
radiation associated with cosmic reionization stunted their formation
\citep[e.g.,][]{bovill2009}.

Though the connection between reionization and truncated star formation in UFDs
appears well-established, the possible contribution of the ancestors of UFDs -- low-mass
star-forming galaxies -- to cosmic reionization is far less
appreciated. \footnote{Especially by Brian.}
In this letter, we address this point by combining the observed stellar fossil
record of UFDs around the MW with stellar population synthesis models in order
to quantify the role of UFDs in reionizing the early Universe. Specifically, we
infer the evolution of the UV luminosities of the lowest-mass UFDs across cosmic
time and explicitly estimate (i) \mmin, the minimum UV luminosity of
star-forming galaxies during the epoch of reionization and (ii) the effects such
systems have on our understanding of the high-redshift ($z \ga 6)$ Universe.
Throughout this paper, we adopt a \citet{planck2016} cosmology.

\section{Methodology}

We combine the star formation histories (SFHs) of UFDs located around the MW
measured from the stellar fossil record with the Flexible Stellar Population
Synthesis code \citep[FSPS;][]{conroy2009, conroy2010a} to compute their
integrated UV luminosities as a function of redshift.  This methodology is detailed
in \citet{weisz2014d} and \citet{boylankolchin2015}.  Here, we summarize the
technique and briefly describe assumptions specific to this analysis.

The SFHs of UFDs have been measured by analyzing deep \hst-based color-magnitude
diagrams (CMDs) for a dozen galaxies located in the immediate vicinity of the MW
\citep[e.g.,][]{brown2014, weisz2014a}. From measurements, we construct a
fiducial SFH of an UFD: 70\% of its total stellar mass
(\mstar($z=0$)\ $\sim5\times10^2$ \msun) forms in a $\sim$100 Myr interval
between $z\sim9-11$, and the remaining 30\% forms over a $\sim$2.5 Gyr period that
ends at $z\sim3$.  This SFH is consistent with observations of Local Group UFDs
\citep[e.g.,][]{brown2014, weisz2014b} and fits well with a scenario in which
the UV background suppresses the accretion of fresh gas onto UFDs, as opposed to
photo-evaporating existing cold gas, allowing for (very) low-level star formation to continue
post-reionization \citep[e.g.,][]{onorbe2015}.

\begin{figure}
\begin{center}
	\includegraphics[width=2.5in]{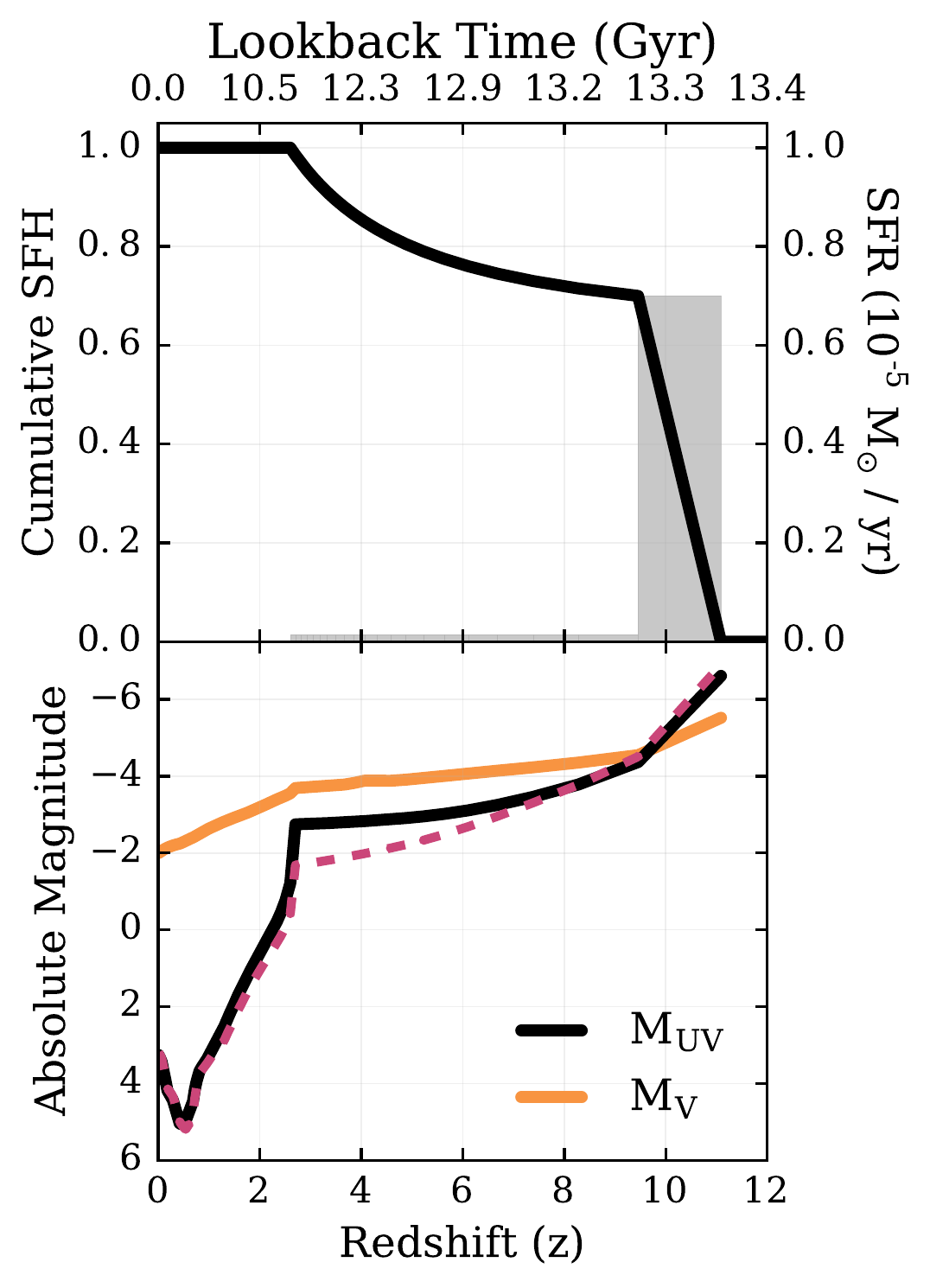}
        \caption{\textit{Top --} The SFH of our fiducial UFD, which is motived
          by the stellar fossil record of UFDs found orbiting the MW.  The grey
          shading indicates the absolute SFR as a function of time and the solid
          black line represents the normalized cumulative SFH.  Patterned after
          analysis of deep HST-based CMDs of UFDs located around the MW, our fiducial UFD
          formed 70\% of between $z\sim9-11$ and 30\% of its stellar mass from
          $z\sim3-9$. The total stellar mass of this system is $\sim5\times10^2$
          \msun, which is comparable to the faintest known galaxies in the Local
          Group. \textit{Bottom --} The evolution of M$_{\rm V}$ (orange) and
          M$_{\rm UV}$ (black) vs. redshift for our fiducial UFD.  The present day
          luminosity, M$_{V}= -2$, is comparable to the faintest known UFDs.
          The magenta dashed line indicates the evolution of \muv\ if UFD formed
          90\% of its stellar mass in the initial event instead of 70\%.}
    \label{fig:ufd_evolution}
\end{center}
\end{figure}

This baseline SFH is plotted in the top panel of Figure \ref{fig:ufd_evolution},
and the bottom panel of the figure shows the corresponding integrated UV (black)
and V-band (orange) flux evolution.  We construct these profiles using FSPS, a
Kroupa IMF \citep{kroupa2001}, the Padova stellar evolution models
\citep{girardi2010}, and a single metallicity of [Fe/H]=-2.0.  We assume no
dust, which is consistent with the observed spectral energy
distributions (SEDs) of the faintest high-redshift galaxies
\citep[e.g.,][]{bouwens2014}. Finally, we normalize the computed fluxes to have
\mv($z=0)=-2$, a value comparable to the faintest known UFDs around the MW
\citep[cf.][]{mcconnachie2012}.

Our fiducial model is designed to broadly reflect the behavior of a typical UFD,
not encompass all possible scenarios.  To illustrate the effects of another
plausible SFH, we consider the case in which 90\% of the stellar mass formed
from $z\sim9-11$. The difference in the UV flux between this alternate SFH and
our default model is no more than 0.5 mag (magenta-dashed line in Figure
\ref{fig:ufd_evolution}).  Beyond varying the SFH, other physical effects
\citep[e.g., interacting binaries, stochastic sampling of the IMF, or various
burst permutations of the SFHs;][]{fumagalli2011, weisz2012a, dominguez2015,
  stanway2016} can affect the UV and ionizing flux output from a galaxy.
However, the amplitude of these effects (typically 1-2 mag) are not sufficiently
large to change the main conclusions of this paper.

\section{Results and Discussion}
\label{sec:results}

Figure \ref{fig:ufd_evolution} illustrates the main result of this paper:
\textit{the stellar fossil record in the Local Group demonstrates that
  star-forming galaxies as faint as \muv\ $=-3$ existed during the epoch of
  reionization ($z\sim6-10$)}. 
Over that period, the UV magnitudes of such
objects initial declines by $\sim3$ mag, as expected from a constant SFH from
$z\sim9-11$.  This period is followed by a gradual decline in integrated UV and
optical flux from $z\sim3-9$, when the SFH is constant but lower by a factor
of $\sim$10 relative to its peak.  Once star formation shuts off at $z\sim3$,
the UV flux falls off precipitously owing to the rapid death of UV-luminous young
stars.  Over the same period, the optical luminosity declines more gradually, as
it originates from stars with a wider range of masses.  Finally, the upturn in UV
luminosity at late times ($z\lesssim0.2$) stems from the onset of blue
horizontal branch stars, which dominate the UV light in the absence of young
stars.  

\begin{figure}
\begin{center}
	\includegraphics[width=2.2in]{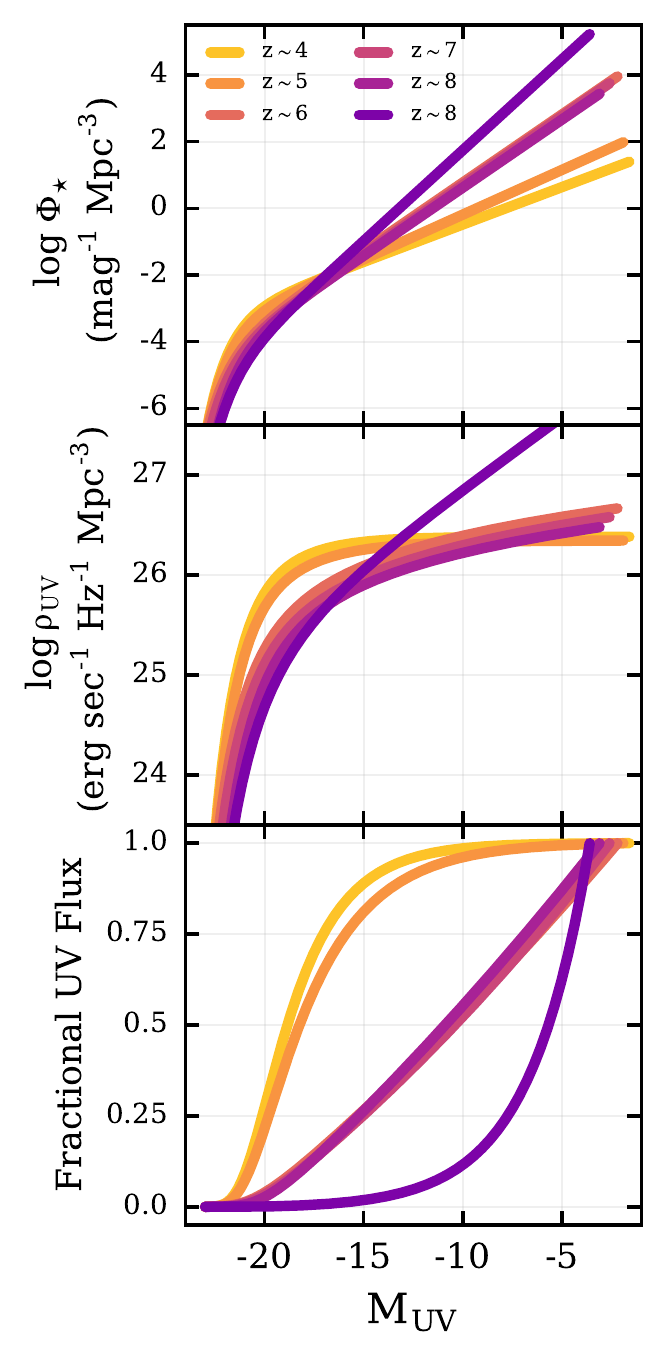}
	\vspace{-10pt}
        \caption{\textit{Top --} Schechter UVLFs from $z\sim4-8$  from
           \citet{finkelstein2015}, along with a $z\sim8$ UVLF from
          \citet{bouwens2015}, based on direct measurements of distant galaxies.  The UVLFS are plotted to \muv\ $\sim-3$, the approximate faint
          magnitude limit based on the stellar fossil record of UFDs found in
          the LG.  Values of the faint-end slopes range from $\alpha\sim-1.6$ at
          $z\sim4$ to $\alpha\sim-2.4$ at $z\sim8$. \textit{Middle --} The integrated UV flux for the UVLFs  in the top panel, plotted as a function of UV magnitude.
           \textit{Bottom --} The
          fraction of UV flux as a function of \muv\ for the Schechter UVLFs
          plotted in the top panel, integrated to
          values 
          plotted in Figure~\ref{fig:ufd_evolution}.  For $z\gtrsim6$, UFDs
          contribute 50-90\% of the total UV flux from galaxies.}
    \label{fig:uvlf}
\end{center}
\end{figure}

\subsection{Assuming Steep Luminosity Functions and Faint Star-Forming Galaxies}
\label{subsec:steep_lfs}
The number density of star-forming galaxies in the early Universe
appears well-described by a Schechter function with a faint-end slope that
steepens with increasing redshift \citep[e.g.,][]{bouwens2015, finkelstein2015}.
In this section, \textit{we briefly explore the consequences of assuming this
  functional form of the UVLF remains valid into the regime of UFDs}. We
consider the alternative -- that the high-$z$ UVLF breaks to a
shallower slope -- in Section~\ref{subsec:broken_lfs}.

\subsubsection{Reionization}

Scenarios for galaxy-driven reionization typically require integration of the
UVLF to a lower luminosity limit of \mmin\ $=-10$ to $-13$ at $z\gtrsim6$
\citep[e.g.,][]{kuhlen2012, robertson2013, robertson2015}.  However, the faint
ancestors of UFDs extend \mmin\ fainter by an additional $\sim$7 magnitudes.
The \muv($z$) values for our fiducial UFD more accurately reflect the true lower
luminosity limit of star-forming galaxies in the early Universe.

Figure \ref{fig:uvlf} shows that extending \mmin\ into the regime of UFDs
(again, assuming the faint-end slope of the LF remains unchanged) results in a
substantial increase in the amount of total UV (and, by extension, ionizing)
flux produced by faint galaxies.  The smallest flux contribution from
UFDs is at low redshifts ($z\sim4-5$), where the faint-end slope is fairly flat
($\alpha\sim-1.5$ to $-1.7$); over 90\% of the flux is
generated by galaxies brighter than \muv$\ =-10$. Faint galaxies make a larger contribution at $z\sim6-7$,  as the faint-end slope is steeper ($\alpha\sim-2$).  Galaxies fainter than current \hst\
blank-field limits (\muv\ $\sim -16$) contribute 80\% of the total UV flux.  Thus, the deepest direct observations of the early Universe only resolve
20\% of the total UV light from galaxies. Further, galaxies fainter than
\muv\ $=-10$, contribute $\sim$50\% of the total
UV flux. Future blank-field observations with \jwst\ are projected to extend to
\muv\ $\sim -14$, which will resolve $\sim$40\% of the total UV flux from galaxies.

Some observation suggest that the faint-end slope may be as steep as $\sim-2.4$ at $z\sim8$ \citep[e.g.,][]{finkelstein2015}, increasing the importance of faint galaxies.
Although uncertainties remain large ($\sim0.4$ dex), we consider two UVLFs at
$z\sim8$, with $\alpha = -2.02$ and $-2.36$ \citep{bouwens2015, finkelstein2015}.  For the shallower slope, the contribution of faint galaxies is identical to the $z\sim6-7$ scenario.
\textit{For the steeper slope, $>$85\% of the instantaneous UV flux comes from
  galaxies fainter than \muv\ $=-10$.} In either case, the deepest blank-field
\hst\ and future \jwst\ observations are only sensitive to at most $\sim10$\% of
the UV light.  This trend may continue at even higher redshifts ($z\sim9-10$),
as current measurements appear to favor faint-end slopes of $\alpha\sim-2.3$
\citep[e.g.,][]{bouwens2016}, albeit with large uncertainties.

Additional UV and ionizing flux from UFDs implies that galaxies could maintain
reionization with smaller escape fractions (\fesc) for ionizing
photons. For example, \citet{robertson2013} assume \fesc\ $=0.2$, $\alpha \sim -2$,
\mmin\ $\sim-10$ for galaxies to maintain reionization.  However, adopting
\fesc\ $=0.1$ and \mmin$=-3.1$ yields the same total ionizing flux.  If the
faint-end slope is as steep as $\alpha\sim-2.4$, then \fesc\ $=0.02$ and
\mmin\ $=-3.1$ would also maintain reionization.  Values of \fesc\ between $\sim2$
and $10$\% appear consistent with several observational and theoretical
results \cite[e.g.,][]{leitet2013, siana2015, ma2015}.

Extra UV flux from UFDs could also imply a higher ionization fraction of the IGM
at times earlier than indicated by \citet{planck2016}. That is, if UFDs formed
early enough and the UVLF was sufficiently steep, then UFDs may have provided
enough flux to initiate reionization earlier than $z\sim10$.  While an
interesting potential conflict, there are still too many uncertainties that
could migiate this concern.  For example, the stellar fossil record of UFDs is
not yet capable of differentiating between star formation that started at
$z\sim10$ and $z\sim12$ (100 Myr in time), which is required to directly compare
measurements from UFDs with constraints from Planck.  For this paper, we have
adopted $z\sim11$ for the start of star formation.  This is in
some tension ($<2\,\sigma$) with Planck constraints on $\tau_{\rm e}$, but
only if the UVLF is very steep at
$z\gtrsim10$.

\subsubsection{Cosmic Star Formation History}

The cosmic SFH is typically derived by adopting a fixed (and fairly bright)
lower bound for integration of the UVLFs at all redshifts (such as L=0.03
\lstar\ or \mmin\ $\sim-17$; e.g., \citealt{madau2014}). Very
faint, star-forming galaxies in the early Universe have a substantial
effect on the cosmic star formation rate (SFR) density at $z\sim 6-8$ in
the steep faint-end slope scenario.

\begin{figure}
  \includegraphics[width=\columnwidth]{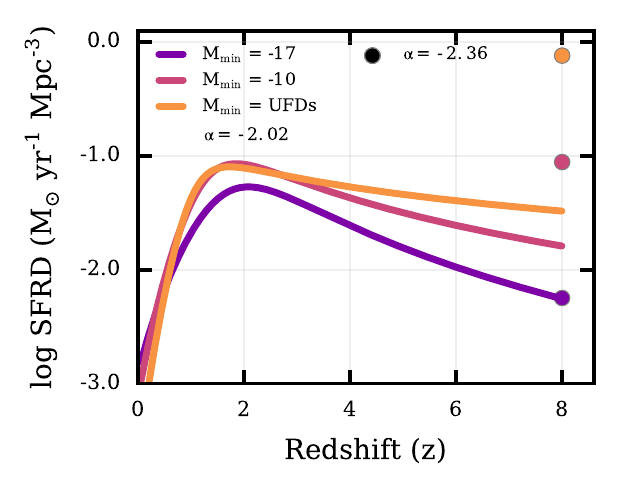}
  \vspace{-20pt}
  \caption{The cosmic SFH, with no correction for dust, computed by integrating
    literature UVLFs at each redshift to different values of \mmin\ and assuming $\alpha=-2.02$ at $z\sim8$.  The solid
    purple line is the fiducial cosmic SFH from \citet{madau2014b}.  The magenta
    line reflects the cosmic SFH when \mmin\ $=-10$, the canonical faint limit for
    galaxies to maintain reionization \citep[e.g.,][]{robertson2015}, while the
    orange line includes the contribution of UFDs.   
    The solid points at $z\sim8$ illustrate the effects of the steeper faint-end
    slope ($\alpha=-2.36$) as reported in \citet{finkelstein2015}.  Including
    UFDs in the calculation of the cosmic SFH can boost the SFR 
    density of the Universe by factors of $\gtrsim4-100$ at $z\gtrsim6$ if the
    UVLF remains steep.  }
    \label{fig:cosmicsfh}
\end{figure}

Figure \ref{fig:cosmicsfh} shows that adopting fainter limits substantially affects the cosmic
SFH for $z\gtrsim6$.  In this Figure, we have fit the functional
form of the cosmic SFH from Equation 15 in \citet{madau2014} to several of the
same UVLFs from the literature used by \citet{madau2014}, adopting the
same UV-to-SFR conversion factor (and making no corrections for
extinction). The various colored lines in Figure \ref{fig:cosmicsfh} show the
cosmic SFH integrated to three values of \mmin: $-17$ (solid purple), $-10$
(solid magenta), and values shown in Figure~\ref{fig:ufd_evolution} for $z\gtrsim3$ and $-10$ otherwise
(solid orange).  The colored points at $z\sim8$ indicate changes to the cosmic
SFH if the steeper faint-end slope of $\alpha=-2.36$ is used
\citep{finkelstein2015}.

Relative to the fiducial assumptions (solid purple), adopting \mmin\ $=-10$
increases the SFR density of the Universe by a factor of $\sim$2-3 for
$z\gtrsim6$.  Adopting values of \mmin\ from the faintest UFDs yields an
increase by factors of $\sim4-6$.  At $z\sim8$, selecting the steeper faint-end
slope ($\alpha=-2.36$) results in substantial changes in the cosmic SFH: an
increase by a factor of $\sim15$ for \mmin\ $=-10$ (magenta point) and $\sim130$
for \mmin\ $=-3.1$ (orange point). Such large increases are also expected at
$z\sim9-10$ if the faint-end slope is steep \citep[$\alpha=-2.3$; e.g.,][]{bouwens2016}.  The large (100x) increase
in SFRs implied by the very steep faint-end slopes at $z\gtrsim6$ may be odds
with the SFR density implied by high-redshift gamma-ray burst observations
\citep[e.g.,][]{kistler2009, kistler2013, chary2016}. However, it remains
unclear if current GRB measurements are unbiased tracers (e.g., that they trace
UFDs) of the total SFR density of the Universe \citep[e.g.,][]{madau2014b}.

\subsubsection{Dark Matter Halo Masses}

Establishing the halo masses of faint galaxies in the early Universe is central
to our understanding of primordial galaxy formation.  Using the stellar-halo mass relation
(SMH) as described in \citet{boylankolchin2014}, we estimate that a
galaxy with \muv$(z=7)=-3.1$ is hosted in a halo with \mhalo$(z=7)\sim10^6$
\msun.  This result is not very sensitive to the adopted SMH relation.

Taken at face value, this halo mass seems in reasonable agreement with predictions for the first mini-halos \citep[e.g.,][]{wise2012}. However, it is also well-below the virial mass ($\sim10^8$ \msun) that
corresponds to the atomic cooling limit ($10^4$ K) at $z \sim 7$.  Galaxy
formation in such low-mass halos is expected to be very inefficient, owing to a
strong reliance on molecular cooling. If the UVLF remains steep into the
regime of UFDs, then there should be a substantial population of UFDs observable
in the Local Group \citep[e.g.,][]{boylankolchin2014, boylankolchin2015}.
However, $\lesssim$10\% of the predicted number of extremely low-mass galaxies
are known to exist in the LG (e.g., see updates to \citealt{mcconnachie2012}).

While environmental effects (e.g., efficient destruction of satellites, dwarf-dwarf mergers, stellar stripping) could alleviate some tension, these mechanisms do not appear efficient enough to resolve the discrepancy \citep[e.g.,][]{kirby2013, deason2014, garrisonkimmel2017}.  Further, the over-abundance of LG dwarfs is not limited to the faintest UFDs, as it is known to persist in the `classical' dwarf regime \citep{boylankolchin2015}, which is thought to be observationally complete \citep[e.g.,][]{koposov2008}.  

Even more challenges arise if (a) the faint-end of the UVLF approaches $\alpha=-2.4$ (i.e., many more faint galaxies) and/or (b) such low-mass galaxies have a low star-formation duty cycle \citep[e.g., $\sim10$\%;][]{wyithe2014}.  In the latter case, 90\% of the halos would be in the `off' state at any given time, and thus have reduced UV luminosities.  Effectively, this results in higher-luminosity galaxies hosted in even lower-mass halos, assuming a canonical abundance matching relationship.

\subsection{Minimizing the Flux Contribution of Ultra-Faint Dwarfs}
\label{subsec:broken_lfs}
Up to this point, we have considered the consequences of extrapolating a
Schechter UVLF into the regime of UFDs.  However, several numerical simulations
of galaxy formation predict a turnover in the UVLF, decreasing the predicted
number of very faint galaxies in the early Universe \citep[e.g.,][]{jaacks2013,
  oshea2015, gnedin2016, liu2016, ocvirk2016, yue2016, finlator2017}. Although details
vary, the common threads among these simulations are that not all low-mass halos
necessarily host a galaxy and that baryonic effects internal to the galaxies
(e.g., supernoave feedback, radiation pressure) can significantly affect galaxy
formation in low-mass halos. Alternately, a turnover in
the high-$z$ UVLF is consistent with models that suppress or eliminate
small-scale structure (e.g., Warm Dark Matter; \citealt{schultz2014,
  dayal2015}).
A broken LF at high-$z$ is also necessary in order to avoid
dramatically overproducing UFDs and even classical dwarfs in the Local Group
\citep{boylankolchin2014, boylankolchin2015}.  

\begin{figure}
\begin{center}
	\includegraphics[width=2.5in]{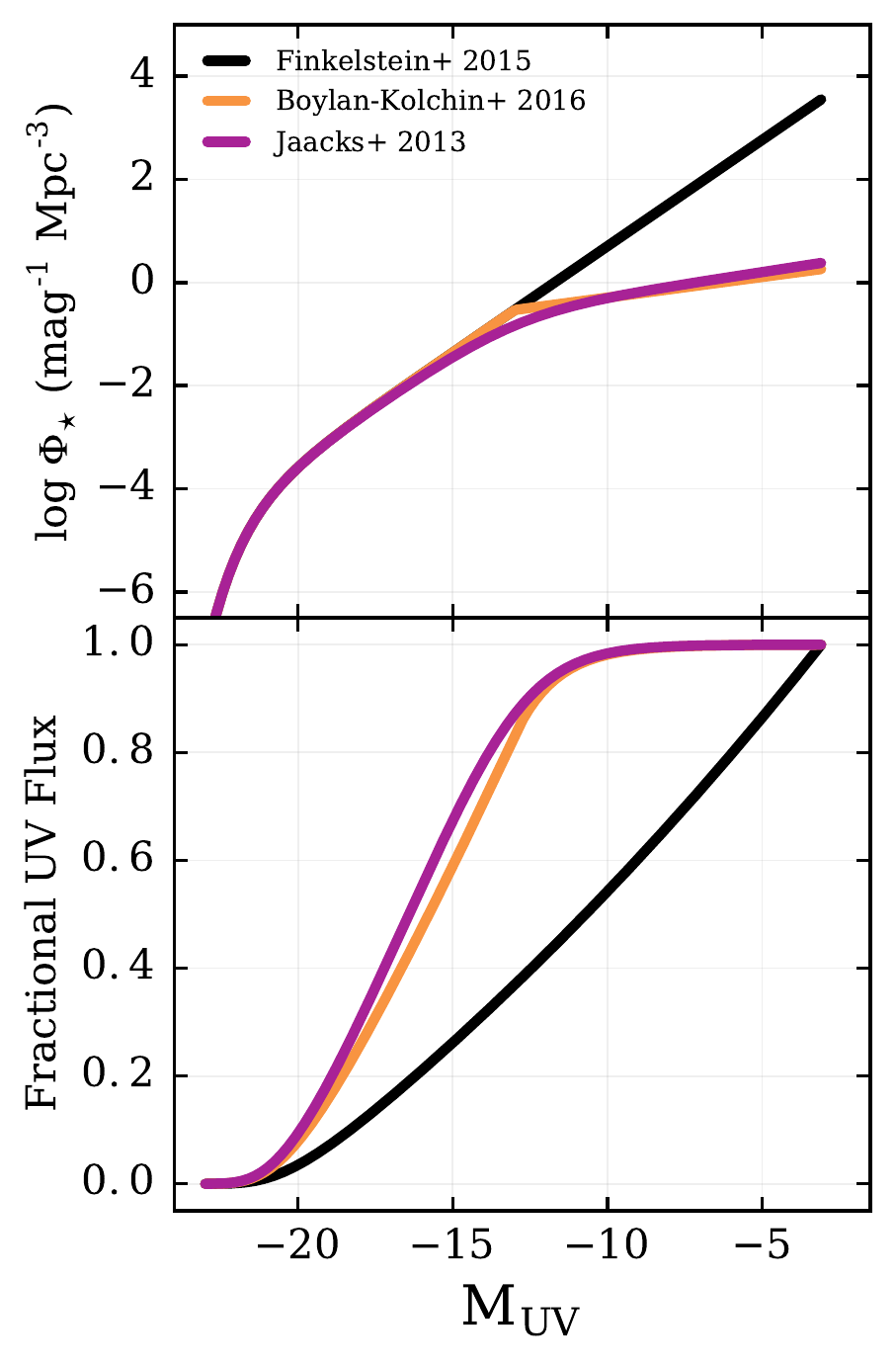}
        \caption{\textit{Top -- } A comparison between the UVLFs at $z\sim7$
          with and without turnovers.  The steep UVLF (black; $\alpha\sim-2$) is
          from \citet{finkelstein2015}, while the other curves show models from
          \citet[orange]{boylankolchin2015} and \citet[magenta]{jaacks2013}.
          \textit{Bottom -- } The
          fraction of UV flux at $z\sim7$ for each of the UVLFs shown in the top
          panel, when integrated to \mmin\ $=-3.1$. A break or turnover in the
          UVLF at \muv\ $=-13$ significantly reduces the flux contribution from
          UFDs.}
    \label{fig:turnover}
\end{center}
\end{figure}

Figure \ref{fig:turnover} illustrates the effects of two plausible UVLF
turnovers.  For simplicity, we focus only the $z=7$ case and adopt analytic UVLFs
that turnover from \citet{jaacks2013} and \citet{boylankolchin2015}.  The
\citet{boylankolchin2015} UVLF has $\alpha\sim-2$ for \muv\ $<-13$ and
$\alpha\sim-1.2$ for \muv\ $\ge-13$ and is designed to reproduce $z=0$ LG dwarf
galaxy counts. For illustrative purposes, we have tuned the UVLF from Equation 1
in \citet{jaacks2013} to match the general shape of \citet{boylankolchin2015}.

The bottom panel of Figure \ref{fig:turnover} shows the cumulative fraction of
UV flux for the UVLFs in the top panel.  The decreased number density for UFDs
reduces their UV flux contribution to $\lesssim10$\%.  We emphasize that this
figure only provides an illustration of the effects of a turnover in the UVLF
and it not necessarily correct in detail.  Further, there is already tentative evidence
that the UVLF remains steep at $z\sim6$ down to at least \muv\ $\sim-12.5$
\citep{livermore2016}, which would be in tension with our selected turnover
parameters (though see \citealt{bouwens2016}).

Beyond matching local galaxy counts and reducing the contribution of UFDs to
reionization, any turnover in the UVLF also affects the SHM relation.  The UVLFs
with a turnover in Figure~\ref{fig:turnover} place \muv$(z=7)=-3.1$ UFDs in
halos with \mhalo$(z=7)=2\times10^8$ \msun~instead of $\sim10^6$ \msun.  Such a
turnover therefore places the faintest galaxies in halos above the atomic
cooling limit, resolving another possible tension implied by a steep UVLF.

\section{Summary}
\label{sec:summary}

By combining the stellar fossil record of known UFDs in the Local Group with
population synthesis modeling, we have demonstrated that star-forming galaxies
with UV luminosities as faint as \muv\ $=-3$ to $-6$ existed during the epoch of
reionization ($z\sim6-10$) must have existed.  We explored the implications of
this result under two possible scenarios, which can be summarized as follows:\\[0.1cm]

\noindent\textbf{If the measured high-redshift UVLFs with steep faint-end slopes are
  valid into the luminosity regime of UFDs:}
 \begin{itemize}
  \item Galaxies fainter than \muv\ $=-10$ contribute $>$50-80\% of the total UV flux from galaxies during reionization.
  \item Galaxies can still power reionization with escape fractions that are $>$50\% lower then currently assumed.
  \item The ancestors of UFDs must have been hosted in halos with \mhalo$(z\sim7)\sim10^6$ \msun, well below the virial mass that corresponds to the atomic cooling limit.
  \item There should be $\sim$10 times more galaxies around the MW with luminosities as bright as Draco (\mv=-8.8).
  \end{itemize}

\noindent\textbf{If the high-$z$ galaxy UVLF turns over at}
  \muv$(z=7)\sim-13$:
\begin{itemize}
\item UFDs do not contribute substantially to reionization at $z \la 8$.
\item There is no number galaxy count tension in the LG.
\item UFDs live in halos more massive than the virial mass associated with the atomic cooling limit.
\end{itemize}

A complete census of Local Group UFDs has the promise to provide a robust
constraint on the faint end of high-$z$ UVLFs.  Even with the current
census, a scenario in which the high-$z$ luminosity function is sharply
truncated at magnitudes significantly brighter than \muv\ $\sim -3$ to $-6$ is
inconsistent with the existence of UFDs in the Local Group. Thus, it appears
unavoidable that the galaxy luminosity function extends, without a sharp
truncation, to luminosities that are $\sim$10,000x fainter than the blank-field
UDF limits.

\section*{Acknowledgements}

The authors thank Ben Johnson for help with correctly translating SFHs to UV
fluxes, and Peter Behroozi, James Bullock, Eliot Quataert, Brian Siana, and Evan Skillman for useful
discussions, particularly regarding the cosmic SFH.  We acknowledge the
extremely valuable discussions at the Near-far workshop in Santa Rosa, CA.
DRW also thanks the SOC of the Galaxy Evolution III conference in Sesto
for the invitation, which led to many new insights. MBK acknowledges
support from the National Science Foundation (grant AST-1517226) and from NASA
through HST theory grants AR-12836, AR-13888, AR-13896, and AR-14282 
awarded by the Space Telescope Science Institute (STScI), which is operated by
the Association of Universities for Research in Astronomy (AURA), Inc., under
NASA contract NAS5-26555. Analysis and plots presented in this paper used
IPython and packages from NumPy, SciPy, and Matplotlib \citep[][]{hunter2007,
  oliphant2007, perez2007, astropy2013}.




\bibliographystyle{mnras}
\bibliography{faintest_galaxies.astroph.bib}




\label{lastpage}
\end{document}